\documentclass[twocolumn,pra]{revtex4}
\usepackage{graphicx}
\usepackage{amssymb}

\begin{document}

\title{The quantum vacuum}

\author{G. S. Paraoanu}
\affiliation{O. V. Lounasmaa Laboratory, School of Science, Aalto University, P.O. Box 15100, FI-00076 AALTO,Finland}

\begin{abstract}
The vacuum is the lowest energy state of a field in a certain region of space. This definition implies that no particles can be present in the vacuum state. In classical physics, the only features of vacuum are those of its geometry. For example, in the general theory of relativity the geometry is a dynamical structure that guides the motion of matter, and, in turn, it is bent and curved by the presence of matter. Other than this, the classical vacuum is a structure void of any physical properties, since classically properties are strictly associated with physical objects such as particles and finite-amplitude fields. The situation is very different in quantum physics. As I will show in this paper, the difference stems from the fact that in quantum physics the properties are not strictly tied to objects. We know for example that physical properties come into existence - as values of observables - only when the object is measured. Thus, quantum physics allows us to detach properties from objects. This has consequences: one does not need pre-existing real objects to create actual properties, and indeed under certain perturbations the quantum vacuum produces observable effects such as energy shifts and creation of particles. An open question is if by necessity the vacuum comes with an embedded geometry, and if it is possible to construct viable physical theories in which geometry is detached from the vacuum.
\end{abstract}

\maketitle

\section{Introduction}

The topic discussed in this paper is the vacuum, an entity that has emerged as an object of intense study in physics. Vacuum is what remains when all the matter, or the particles corresponding to all the known fields, are removed from a region of space. The philosophical question that this paper aims at addressing is: in what sense can this remaining entity be said to exist? Does it have any properties - and how can it have properties if nothing is there? If it doesn't have any properties, how can it be described at all - there seems to be nothing to talk about. I will argue that quantum theory offers a radical departure from the classical concept of property as an attribute of an already-existing particle or field. This has testable consequences: measurable effects ({\it e.g.} energy shifts, radiation, effects on phase transitions) can be created in a certain region of space without the need of physical objects as carriers of those properties in that region.

In the following, I will use the word ``real'' or ``actual'' as referring to classical events - events that leave, somewhere in nature or in the laboratory, a classical record or trace. ``Classical'' is what both the theory of relativity and quantum mechanics agree upon. Examples of real entities are the results (values) of a measurement in quantum physics, the clicks of particle detectors, the bit of information (either 0 or 1) stored or recorded in a macroscopic register, and the events from the theory of relativity. I take for granted that the classical world is real. For everything else - mathematical constructs, quantum states, structures {\it etc.}, I will use ``to exist'' and ``to be'' in a rather generic way - otherwise I wouldn't know how to refer to these entities. Also I will use of the words ``property'' and ``structure'' as referring both to actual and potential properties and structures. Real (actual) entities certainly have real (actual) properties, while structures can be regarded formally as sets of properties with certain relations between them. For example, the momentum and the position at a certain time of a classical particle are real properties, while the Lagrangian, the law of motion, the Poisson bracket, {\it etc.} are structures. The difference is that the former are the result of a direct measurement, independent of the law of motion, while the latter describe a specific law of dynamical evolution satisfied by these properties. But also virtual or potential properties, such as the unmeasured value of one observable when the system is in an eigenstate of the canonically conjugate observable, can form structures. Examples of  such structures are the modes of the electromagnetic field in a cavity, the geometry, or the non-commutativity of operators in the Hilbert space. As we will see, the main conclusion of this paper is that the quantum vacuum is an entity endowed with plenty of structure, which lies beneath the existential level of ``real'' matter. Its ontological status as the seed of possibilities is derived from quantum physics. I give first a brief sketch of the historical development leading to the present notion of quantum vacuum, then I present a few interesting open problems and connections between several lines of investigation of this concept, both in physics and philosophy.

\section{Brief historical interlude: the development of the concept of vacuum up to the quantum era}

Philosophical reflection about vacuum is as old as philosophy itself. The Greek atomists Leucippus and Democritus were the first to be worried whether the vacuum is a well-defined concept or not. For them this was important, because, after all, what is left between the atoms must be the void. If atoms are to be taken as constituents of the world, then so must be the void. Thus the atomists clearly saw that both the full (the being) and the empty (the non-being) have to be postulated as the primordial elements.

Aristotle, following Plato's thought, devised a number of rather ingenious arguments against the existence of vacuum \cite{grant}. Some of these arguments have to do with the difficulty of making sense of motion in vacuum, as one would need some reference points with respect to which to describe changes of the position. But in vacuum all points would be equivalent, therefore, worries Aristotle, motion cannot be defined. Moreover, motion in vacuum, if vacuum exists, should continue forever, in flagrant contradiction with Aristotle's own physics that assessed that motion is due to things aiming at reaching their natural place. Besides these physics - based arguments, Aristotle formulated a ``logical'' argument against the existence of vacuum. Suppose one removes a body from the place it occupies in space. If we were to attribute any reality to the emptiness left behind, we would need to refer to it as a body with the characteristics of existence (being). Imagine now that we put back the body where it was. We are now left with two bodies co-existing in exactly the same region of space. In this case, Aristotle thinks, the vacuum would need to have rather miraculous properties. It should be more like a fluid that perfectly penetrates the initial body, filling exactly the same amount of space.

The Middle Ages did not bring up any significant deviation from Aristotle's arguments. Nature's abhorrence of vacuum ({\it horror vacui}) was accepted by most thinkers. But what did eventually turn the tables around in favor of vacuum was the experiment: in the XVIIth century a series of experiments due to Torricelli, Pascal, and von Guericke demonstrated that removing the air from an enclosure is technically possible, and from that moment on the vacuum became a legitimate object of study for science. Its ontological status remained however unclear and would change several times during the next centuries. As we will see below, it has remained, until nowadays, tied with the concept of space, and as a result it would go through the reformulations imposed by the Newtonian mechanics, by the theory of relativity, and by the quantum physics.

In Decartes's philosophy the refutation of the reality of absolute space is mostly based on the association between extension and bodies. If bodies are removed, then one cannot talk about extension anymore - thus absolute space is absurd. What we call space is then an ensemble of contiguities: the location of a body is a collection of relations between the body and those immediately contiguous to it. Motion is simply a change in these contiguity relations \cite{Descartes}.

Against this type of relationist thinking, due to Descartes and to Leibniz as well, Newton exposes his conception of absolute space and time in the famous {\it Scholium} of {\it Principia} \cite{Newton},

\vspace{2mm}
``Only I must observe, that the common people conceive those quantities under no other notions but from the relation they bear to sensible objects. And thence arise certain prejudices, for the removing of which it will be convenient to distinguish them into absolute and relative, true and apparent, mathematical and common.''
\vspace{2mm}

 Following the great success of Newtonian mechanics, space had become established as the universal receptacle of objects. However, the  ``common people'' (``{\it vulgus}'' in the original) eventually had their way: the corpuscular view of light advocated by Newton had to yield to the wave view of his contemporary Huygens. Later in the XIX'th century the wave theory of light would get experimental confirmation through the work of Young and Fresnel, and will be put on solid mathematical grounds by Maxwell. But light needed a medium into which to propagate as a wave, so it was conjectured that such a medium called ``ether'', filling the absolute space, would exist. At the end of the XIXth century, the experiments of Michelson and Morley showed however that there is no motion with respect to the ether.

 Finally, the theory of relativity of Einstein made redundant the concept of ether and that of absolute time and space. The conceptual pendulum swang back to the relationists' side \cite{saundersbrown}. The special theory of relativity introduced the idea that length and time intervals are not absolute quantities, but, instead, they depend on the state of motion of the observer. The Lorentz transformation and the negative result of the Michelson and Morley experiment were explained as a natural consequence of the postulates of relativity. From now on, spacetime has become defined only in relation to a reference frame, with each object dragging with it its own spacetime as it moves. It is the concept of motion that forces us to attach different vacua to objects moving with respect to each other. Einstein explains it with exquisite clarity \cite{Einstein},

 \vspace{2mm}
 ``When a smaller box {\sl s} is situated, relatively at rest, inside the hollow space of a larger box {\sl S}, then the hollow space of {\sl s} is a part of the hollow space of {\sl S}, and the same `space', which contains both of them, belongs to each of the boxes. When {\sl s} is in motion with respect to {\sl S}, however, the concept is less simple. One is then inclined to think that {\sl s} encloses always the same space, but a variable part of the space {\sl S}. It then becomes necessary to apportion to each box its particular space, not thought of as bounded, and to assume that these two spaces are in motion with respect to each other.''
 \vspace{2mm}

 Finally, general relativity puts gravitation and noninertial motion into this picture. In the theory of general relativity the coordinates (space and time) are even more devoid of any physical meaning than in special relativity.
 The metric is itself a solution of Einstein's equations - if this solution exists, space-time can be rightfully said to exists. If it does not, such as in the singularities of black holes or in the Big Bang, spacetime does not have any meaning. If one somehow removes the metric, as given by solving Einstein's equations, what is left is not the absolute flat spacetime of Newton - nothing is left.

In Einstein's words \cite{Einstein},

\vspace{2mm}
``There is no such thing as an empty space, {\it i.e.}
a space without field. Space-time does not claim existence on its own,
but only as a structural quality of the field. Thus Descartes was not so far from the truth when he believed he must exclude the existence of an empty space.''
\vspace{2mm}

Here the word field refers to the gravitational field, which in the general theory of relativity can be seen, so to say, almost co-substantial with the metric $g_{\mu\nu}$ (that is why it is also called a metric field). This is, in brief, the great conceptual shift introduced by the general theory of relativity: that spacetime is a field with a dynamics of its own, as determined by the configuration of matter, and not just a fixed background attached to each reference frame, as in the special theory of relativity. In modern mathematical parlance, we say that general relativity is a background-independent theory (due to diffeomorfism invariance), meaning that the theory is not built on a fixed spacetime geometry that exists behind the scenes, unaffected by matter. Einstein's equations tell us explicitly that there can be no such background that is left unbent by the action of matter.

\section{The architecture of the quantum vacuum}

Three major philosophical assumptions about properties can be associated with the Newtonian world-view of the world.

{\bf [A1]} Properties are tied to physical objects (particles or non-zero fields).

{\bf [A2]} Space is distinct from and exists independently of the objects (carrying properties) one chooses to populate it with. The same is true for time. Space-time is the immense theater stage where physical processes unfold, the canvas where each dot is an event. One has, in principle, access to any of these points.

{\bf [A3]} True randomness does not exist. The observed randomness of the properties of a system is simply a result of our lack of knowledge and imperfect control over the experiment.

The rise of electrodynamics in the mid-19th century did not change much [A1]. It only added fields, mostly through the work of Faraday, as legitimate carriers of properties. Neither did Boltzmann's statistical mechanics present a challenge to [A3], since there the perceived randomness was presumed to be an effect of the motion and collisions of many particles.
It was the general theory of relativity that changed [A2] to a large extend: the theory suggests that spacetime itself can be bent due to the presence of matter. A distribution of matter allows one to calculate the metric of spacetime. But accessing any of the points of spacetime is no longer taken for granted - there can exist points where the theory predicts singularities, event horizons prevent the transfer of information from the inside of the region of space which they enclose, and so on.

On this issue, the theory of relativity is not as radical as one can be. As we have seen, for the thinkers before Newton the connection between objects and spacetime was even tighter. Spacetime might not mean anything in the absence of objects. However, even in this conceptual frame it still makes sense to ask what  happens when we attempt to remove all the objects from a certain region. There are three possible answers: the first, that the problem is logically ill-defined. This seems to be what Aristotle preferred to believe. Another possibility is to view the objects and their associated spacetime as analogous to fluids: attempting to remove a part of the fluid is hopeless because it will be immediately replaced by another part of the fluid. Finally, the third view could be called a  ``ceramical'' view of spacetime: much like the tiles in a glass mosaic, any attempt to remove one part of the drawing results in the  breaking of the glass, extracting objects from the spacetime could result simply in some type of nothingness. In this case, because spacetime is so rigidly attached to the physical objects, it makes sense to wonder if a spacetime structure is useful at all or it is just redundant. This type of conceptual structure might not allow to construct a physical theory in the usual sense: as a story that unfolds in spacetime - simply because there is no spacetime, or it is not clearly distinct from the objects themselves.

But there is another way out. Quantum physics offers a completely different perspective that completely changes [A1] and [A3], and softens the alternatives to [A2] by introducing more conceptual structure. The result is essentially a probabilistic theory in which the evolution is not applied directly to probabilities, but to probability amplitudes \cite{aaronson}. This automatically means that what evolves are not the properties of the objects, but the possibilities of the objects having certain properties. These properties become actualized (real) only after a measurement of the corresponding observable. The assumption [A2] is to some extend left unscathed: spacetime exists as an independent entity, but it acquires more structure beyond just geometry.

Before proceeding further, it is worth stressing out that so far \cite{amps} there is no prediction of quantum physics that contradicts the general theory of relativity, or the other way around. Of course, we do not know the limits of these theories, and one may reasonably suspect for instance that quantum mechanics will forbid the point-like singularities of general relativity. Still, the fact that the known domain of applicability of both of these theories is so vast - from elementary particles to structures of the size of galaxies - and yet no contradictory result has been obtained is astonishing, especially when one looks at how different are the concepts and assumptions the two theories operate with. Even the combination of the special theory of relativity with quantum physics in the form of relativistic quantum field theory, producing the very successful predictions of the Standard Model of particle physics, is not an easy conceptual marriage \cite{wigner}. This situation is rather unique in the history of science, and brings a novel twist to the discussion on falsifiability \cite{popper}, paradigm shifts \cite{kuhn}, confirmation, {\it etc.}, which would be worth investigating further in the philosophy of science.

Another remark is that the assumptions [A1]-[A3] are not exhaustive. I left aside for example the very important supposition that interaction is strictly local (there is no action-at-a-distance), a feature which is essential when considering the dynamics of systems. This assumption is maintained in standard quantum field theory - when writing the interaction Hamiltonian between two fields, it is taken for granted that one field couples only to the other field defined at the same point in spacetime. This, however, does not make quantum physics local in the classical sense, because once they have interacted the particles (or fields) can be separated in space and some of the properties that one ascribes to them via measurements cannot result from local probability distributions. This type of quantum non-locality, as famously put first in evidence by Einstein, Podolsky and Rosen, is best expressed by Bell inequalities, but, interestingly, it can also be put in evidence as a purely logical contradiction \cite{paraoanu}.

\subsection{The ontological status of the quantum vacuum}

In quantum physics, vacuum is defined as the ground state of a quantum field. It is a state of minimum energy, corresponding to zero particles. Note that this definition of vacuum uses already the conceptual and formal machinery of quantum field theory. It is justifiable to ask weather it is possible to give a more theory-independent definition with lesser theoretical load. In this situation vacuum would be an entity which is explained - not just defined within and then explored - by quantum field theory. For example, one could attempt an operational definition of vacuum as the state in which no particles are detected. But then we have to specify how to detect the particles, with what efficiency, {\it etc.}, that is, we need a model for the particle detector. Such a model, known as the Unruh-DeWitt detector, is constructed however from within quantum field theory. Therefore nothing seems to be gained in explanatory power by an operational definition.

The vacuum is simply a special state of the quantum field -  implying that quantum physics allows the return of the concept of ether, although in a rather weaker, modified form. This new ether - the quantum vacuum - does not contradict the special theory of relativity because the vacuum of the known fields are constructed to be Lorentz-invariant. In some sense, each particle in motion carries with it its own ether, thus Lorentz transformations act in the same way on the vacuum and on the particle itself. Otherwise, the vacuum state is not that different from any other wavefunction in the Hilbert space. Attaching probability amplitudes to the ground state is allowed to the same degree as attaching probability amplitudes to any other state with nonzero number of particles. In particular, one expects to be able to generate a real property - a value for an observable - in the same way as for any other state: by perturbation, evolution, and measurement. The picture that quantum field theory provides is that both  particles and vacuum are now constructed from the same ``substance'', namely the quantum states of the fields at each point (or, equivalently, that of the modes). What we used to call matter is just another quantum state, and so is the absence of matter - there is no underlying substance that makes up particles as opposed to the absence of this substance when particles are not present. One could even turn around the tables and say that everything is made of vacuum - indeed, the vacuum is just one special combination of states of the quantum field, and so are the particles. In this way, the difference between the two worldviews, the one where everything is a plenum and vacuum does not exist, and the other where the world is empty space (nonbeing) filled with entities that truly have the attribute of being, is completely dissolved. Quantum physics essentially tells us that there is a third option, in which these two pictures of the world are just two complementary aspects. In quantum physics the objects inhabit at the same time the world of the continuum and that of the discrete.

Incidentally, the discussion above has implications for the concept of individuality, a pivotal one both in philosophy and in statistical physics. Two objects are distinguishable if there is at least one property which can be used to make the difference between them. In the classical world, finding this property is not difficult, because any two objects have a large amount of properties that can be analyzed to find a different one. To establish if a painting is fake or it is the original is only a matter of practical difficulty. But, because in quantum field theory objects are only combinations of modes, with no additional properties, it means that one can have objects which cannot be distinguished one from each other even in principle. For example, two electrons are perfectly identical. To use a well-known Aristotelian distinction, they have no accidental properties, they are truly made of the same essence. A very important related problem is that of the distinguishability of non-orthogonal states, which has attracted a lot of attention in quantum information.

Another spectacular application of the idea that properties are detached from objects is quantum computing. Unlike in classical computing, quantum processors do not need to use objects (for example memory elements) as physical support for each of the intermediate result of a calculation \cite{persp}. The re-attachment of properties in the form of the result of a calculation is done only at the end of a series of unitary operations, when the registers are measured.

To see in a simple way why quantum physics requires a re-evaluation of the concept of emptiness the following qualitative argument is useful: the Heisenberg uncertainty principle shows that, if a state has a well-defined number of particles (zero) the phase of the corresponding field cannot be well-defined. Thus, quantum fluctuations of the phase appear as an immediate consequence of the very definition of emptiness.

Another argument can be put forward: the classical concept of emptiness assumes the separability of space in distinct volumes. Indeed, to be able to say that nothing exists in a region of space, we implicitly assume that it is possible to delimitate that region of space from the rest of the world. We do this by surrounding it with walls  of some sort. In particular, the thickness of the walls is irrelevant in the classical picture, and, as long as the particles do not have enough energy to penetrate the wall, all that matters is the volume cut out from space. Yet, quantum physics  teaches us that, due to the phenomenon of tunneling, this is only possible to some extent - there is, in reality, a non-zero probability for a particle to go through the walls even if classically they are prohibited to do so because they do not have enough energy. This already suggests that, even if we start with zero particles in that region, there is no guarantee that the number of particles is conserved if {\it e.g.} we change the shape of the enclosure by moving the walls. This is  precisely what happens in the case of the dynamical Casimir effect, as described below. Another consequence, which I will not discuss here, is the existence  of entanglement between different regions of space in the vacuum state, a somewhat unexpected effect since the concept of entanglement is usually discussed for particles. There is yet another point of view that illustrates that in quantum physics the idea of delimitating a region of space, and taking the particles out of it, is tricky. The very concept of a particle is not a local one in quantum field theory \cite{rovelli}, and defining the number of particle operator in a region of space is not trivial \cite{redhead}. Particles are extended objects but the operation of removing  them is by necessity local - thus when abstractly separating an empty volume of space one needs further care to ensure that no particle leaks in.

All these demonstrate that in quantum field theory the vacuum state is not just an inert background in which fields propagate, but a dynamic entity containing the seeds of multiple possibilities, which are actualized once the vacuum is disturbed in specific ways. This leads to real effects, some of which are discussed in the next subsection: vacuum fluctuations result in shifts in the energy level of electrons (Lamb shift), fast changes in the boundary conditions or in the metric produce particles (dynamical Casimir effect), and accelerated motion and gravitation can create thermal radiation (Unruh and Hawking effects).

\subsection{Observable effects due to the quantum vacuum}

There are several field-theoretical and many-body effects associated with the existence of vacuum fluctuations \cite{sciama,milonni}.

Measurements showing conclusively that differences in vacuum energies have observable effects provided some of the earliest experimental confirmations of quantum physics. For example, one possibility is to measure the vibrational spectra of molecules and to search for isotope effects (a change in the mass of a nuclei will change the zero-point energy, thus the transition frequencies). The first observation of this effect was done by  Mulliken in 1925, using boron monoxide. Since then, the vacuum state has played an important role in countless other experiments. For example, in X-ray scattering on solids, it was shown that the zero-point fluctuations of the phonons produces an additional scattering on top of that due to thermal fluctuations. Other examples are the Lamb shift between the energies of the s and p levels in the hydrogen atom, and the fact that liquid helium does not become solid at normal atmospheric pressure even near zero temperature - the vacuum fluctuations prevent the atoms of coming close enough so that solidification can occur. In nuclear physics, a related problem is that of a fundamental limit of the size of nuclei \cite{alpha}. As the charge number Z increases beyond approximately $1/\alpha$ (where $\alpha$ is the fine structure constant), the electric fields near the nucleus produce vacuum instability \cite{nuclear}, and particle-antiparticle pairs are generated from vacuum due to the Schwinger effect.

The dynamical Casimir effect was predicted theoretically in 1970 \cite{Moore} and has been recently observed in two experiments. The first one uses a SQUID terminating a coplanar waveguide \cite{ChalmersNature}, creating a fast-moving boundary condition. The other experiment employs an array of SQUIDs, effectively realizing a material with a fast-tunable index of refraction embedded in a cavity \cite{PNAS}. When the boundary condition (in the first setup) or the index of refraction (in the second setup) changes fast enough, one observes real photons emerging from the circuit, even if the system was initially in the vacuum state. Quantum superfluids offer also a rich system to observe vacuum effects: such experiments have been discussed in superfluid He \cite{Volovik}, and recently a thermal analog of the dynamical Casimir effect has been reported in a Bose-Einstein condensate \cite{westbrook}.

In order to understand conceptually the dynamical Casimir effect let us go back to Einstein's {\it gedankenexperiment} with the two boxes $S$ and $s$, as presented in the second section. Einstein realized that motion imposes on us the concept of  a relativistic, frame-dependent space. This relative space is dragged along by the box (or frame) as it moves. As a result, space is not just a kind of fixed canvas onto which we draw reference frames, but, instead, it is defined by and anchored into the reference frame.  With this, we are now ready to push Einstein's thought one step further. Because space is an entity effectively created by some enclosure, this implies that deforming the corresponding box or boundary condition might have an effect on the space inside.  For example, we can compress and expand the space itself by operating the box as a piston in a cylinder. The result turns out to the creation of real particles. Einstein would have been amazed: quantum physics brings his own view on space to a very unexpected  consequence!

Finally, motion itself has an effect on the vacuum.  Let us look again back at Einstein's boxes $s$ and $S$. Each of them carry their own vacuum. As they move one with respect to each other, is the vacuum of $S$ also seen as a vacuum by $s$? The principle of relativity guarantees that
no phenomenon exists allowing to distinguish the vacua of the two inertial systems, but for non-inertial motion it does not put any restriction. It turns out that if $s$ is moved with respect to $S$ at a constant acceleration, $s$ experiences a thermal background (an environment containing particles in thermal equilibrium). This is the Unruh effect \cite{unruh}. Now, by the principle of equivalence, gravitation is equivalent to acceleration, so one expects a similar effect to occur in gravitational fields. This is the famous Hawking effect \cite{hawking}, consisting in emission of radiation at the event horizon of a black hole.

\subsection{Where do properties come from}

We now go back to the main theme of this paper: what is the origin of the properties of physical objects? As we have seen, we have to enlarge the category of entities where properties can originate from, by including the quantum vacuum. To make the  difference more clear, suppose that we have a region of space emptied of matter and fields. Nothing real, in the sense defined in the introduction, is there. Classically, the only  way to create a property inside that region is to bring in from outside an object carrying that specific property. Note that this simple thought experiment relies on all of the assumptions [A1], [A2], and [A3] listed above. These are not trivial assumptions - although they look very innocuous, it is by no means obvious that nature should obey them. In this sense, Netwonian physics appears as a strongly coerced theory, while relativity and  quantum physics introduce different relaxations of these assumptions.
Firstly, Newtonian physics needs to have the concept of space as in [A2], existing independently of objects and with all the points easily accessed. General relativity shows that this does not happen if  the object carrying  the desired property is too massive or if we insist of making it as much as point-like - squeezing too much energy into too little space could result in the formation of a black hole. Secondly, if [A1] and [A3] is not satisfied, then properties could appear spontaneously in vacuum, as they do not require either a real object to be  attached to or a causal chain of events that would produce them.

The experiments on generation of particles from the quantum vacuum mentioned above (dynamical Casimir effect) show that there exists another way of generating properties. Note that these experiments still use the classical concept of spacetime background as in [A2], but to explain them one needs to alter dramatically [A1] and [A3] to  accommodate the quantum-mechanical account of randomness (there exists pure randomness) and properties (properties are not intrinsically attached to objects, but are created contextually, as shown by the Kochen-Specker theorem). Because in quantum field theory the vacuum has a structure, properties can be generated at a certain point by changes of this structure, and not just by bringing them in from somewhere else. As mentioned above already, one cannot do this classically: if a property were to appear at some point in space, then classical physics would tell us that, according to [A1] there must be a real object that carries this property, and according to [A2] there must be a causal story, enfolding in the region of space-time under consideration, which one must discover in order to have a complete description of the phenomenon. To clarify this point, I can make an analogy with the chairs for the public in a concert hall. The arrangement of chairs in rows and the numbering of the chairs in each row, the association of higher prices to better seats {\it etc.}, provides a structure for the probability distribution of spectators.
For example, if one tries to buy a ticket, the options are limited by the total number of seats, by the number of already-reserved seats, and by the budget of that person.
The spectators are here the properties: they might buy a certain seat and show up to the concert -  or not.   However, to create this arrangement of seats in the concert hall one needs to bring in the chairs from outside: there must be some energy and mass to support this structure, and this energy and mass can be recovered if for example the concert hall is renovated and the chairs are removed. This situation is in contrast to the quantum vacuum, where the structure exists as such, ready to acquire real properties, without being constructed beforehand
by energy or mass previously brought in from elsewhere. By definition, the vacuum is the ground state, therefore (unless the system is metastable) there is no other lower-energy state where the system would go to if one attempts to extract energy from it. This feature has experimental applications, for example to verifying that systems such as nanomechanical oscillators have reached the ground state \cite{nanomech}. Note that in the case of the dynamical Casimir experiments mentioned above, the energy of the particles comes from the pump in a two-photon spontaneous downconversion process: the vacuum only provides a structure for this process to occur, and it is not the case that the vacuum energy is converted into photons. In general, deforming, shearing, modifying boundary conditions, and changing the index of refraction of the vacuum results in energy exchange - for example, in the static Casimir effect it costs energy to pull apart the two plates. The quantum vacuum behaves, from this point of view, almost as a real material. Clearly, the ontological status of an entity that is not made of real particles but reacts to external actions does not fall straight into any of the standard philosophical categories of being/non-being.

\section{An emptiness full of unknowns}

A significant number of important open problems in physics are connected to the concept of vacuum. I will briefly discuss here a few of them.

\subsection{What lies beneath the continuous spacetime manifold}

If the quantum vacuum displays features that make it resemble a material, albeit a really special one, we can immediately ask: then what is this material made of? Is it a continuum, or are the ``atoms'' of vacuum? Is vacuum the primordial substance of which everything is made of? Such questions lead us to the very edge of our knowledge. To make these big questions more  understandable, we can start by decoupling the concept of vacuum from that of spacetime.

As we have seen, the concept of vacuum as accepted and used in standard quantum field theory is tied with that of spacetime. This is important for the theory of quantum fields, because it leads to observable effects. It is the variation of geometry, either as a change in boundary conditions \cite{ChalmersNature} or as a change in the speed of light (and therefore the metric) \cite{PNAS} which is responsible for the creation of particles. Now, one can legitimately go further and ask: which one is the fundamental ``substance'', the space-time or the vacuum? Is the geometry fundamental in any way, or it is just a property of the empty space emerging from a deeper structure?

These questions force us to go back to reexamining the most basic conceptual cornerstones of our physical theories.
That geometry and substance can be separated is of course not anything new for philosophers.
Aristotle's distinction between form and matter is one example. For Aristotle the ``essence'' becomes a true reality only when embodied in a form. Otherwise it is just a substratum of potentialities, somewhat similar to what quantum physics suggests. Immanuel Kant was even more radical: the forms, or in general the structures that we think of as either existing in or as being abstracted from the realm of independently-existing reality (the thing-in-itself or the {\it noumena}) are actually innate categories of the mind, preconditions that make possible our experience of  reality as {\it phenomena}. Structures such as space and time, causality,  {\it etc.} are {\it a priori} forms of intuition - thus by nature very different from anything from the outside reality, and they are
used to formulate synthetic {\it a priori} judgments. But almost everything that was discovered
in modern physics is at odds with Kant's view \cite{heisenberg}. In modern philosophy perhaps Whitehead's process metaphysics \cite{whitehead} provides the closest framework for formulating these problems. For Whitehead, potentialities are continuous, while the actualizations are discrete, much like in the quantum theory the unitary evolution is continuous, while the measurement is non-unitary and in some sense ``discrete''. An important concept is the ``extensive continuum'', defined as a ``relational complex'' containing all the possibilities of objectification. This continuum also contains the potentiality for division; this potentiality is effected in what Whitehead calls ``actual entities (occasions)'' - the basic blocks of his cosmology. For the pragmatic physicist, since the extensive continuum provides the space of possibilities from which the actual entities arise, it is tempting to identify it with the quantum vacuum \cite{hattich}. The actual entities are then assimilated with events in spacetime, as resulting from a quantum measurement, or simply with particles. The following caveat is however due: Whitehead's extensive continuum is also devoid of geometrical content, while the quantum vacuum normally carries information about the geometry, be it flat or curved.

It is reasonable to expect that the continuous differentiable manifold that we use as spacetime in physics (and experience in our daily life) is a coarse-grained manifestation of a deeper reality, perhaps also of quantum (probabilistic) nature. This search for the underlying structure of spacetime is part of the wider effort of bringing together quantum physics and the theory of gravitation under the same conceptual umbrella. From various theoretical considerations, it is inferred that this unification should account for physics at the incredibly small scale set by the Planck length, $10^{-35}$m, where the effects of gravitation and quantum physics would be comparable. What happens below this scale, which concepts will survive in the new description of the world, is not known. An important point is that, in order to incorporate the main conceptual innovation of general relativity, the theory should be background-independent. This contrasts with the case of the other fields (electromagnetic, Dirac, {\it etc.}) that live in the classical background provided by gravitation.

The problem with quantizing gravitation is - if we believe that the general theory of relativity holds in the regime where quantum effects of gravitation would appear, that is, beyond the Planck scale  - that there is no underlying background on which the gravitational field lives. There are several suggestions and models for a ``pre-geometry'' (a term introduced by Wheeler) that are currently actively investigated (see {\it e.g.} \cite{pregeometry} for a non-technical review).
This is a question of ongoing investigation and debate, and several research programs in quantum gravity
 (loops, spinfoams, noncommutative geometry, dynamical triangulations, {\it etc.}) have proposed different lines of attack \cite{boi}. Spacetime would then be an emergent entity, an approximation valid only at scales much larger than the Planck length.

Incidentally, nothing guarantees that background-independence itself is a fundamental concept that will survive in the new theory. For example, string theory is an approach to unifying the Standard Model of particle physics with gravitation which uses quantization in a fixed (non-dynamic) background. In string theory, gravitation is just another force, with the graviton (zero mass and spin 2) obtained as one of the string modes in the perturbative expansion. A background-independent formulation of string theory would be a great achievement, but so far it is not known if it can be achieved.

Models of emergent spacetimes can be constructed by analogy with the low-energy models used in condensed-matter physics \cite{bain}. One recent particularly simple to understand such construction is the quantum graphity model of Markopoulou and collaborators \cite{markopoulou}, a model inspired from loop quantum gravity. In this model the geometry emerges from a probabilistic structure which is itself of quantum-mechanical nature: geometrical relations are given by the links between the nodes  of a graph, and these links are created and annihilated by standard quantum-mechanical creation and annihilation operators. Two nodes are in a relation of spatial vicinity only if the link between them is in the state ``connected'', as resulting from the action of the creation operator on the vacuum. Note that the graph does not live in spacetime: it is an abstract lattice describing connection relationships between nodes. The geometry is emergent as the overall connectivity of the graph. The concept of proximity is therefore probabilistic (in the sense of quantum mechanics) and it allows for states that are quantum-mechanical superpositions of connected or disconnected links, yielding also superpositions of geometries.

\subsection{Time, gravitation, energy, and the origin of the Universe}

The relationship between the quantum vacuum and other fundamental concepts in physics such as time, gravitation, and energy is not easy to pin down, but some of these connections are intriguing.

Time is one of the most difficult concepts in physics. It enters in the equations in a rather artificial way - as an external parameter. Although strictly speaking time is a quantity that we measure, it is not possible in quantum physics to define a time-observable in the same way as for the other quantities that we measure (position, momentum, {\it etc.}). The intuition that we have about time is that of a uniform flow, as suggested by the regular ticks of clocks. Time flows undisturbed by  the variety of events that may occur in an irregular pattern in the world. Similarly, the quantum vacuum is the most regular state one can think of. For example a persistent superconducting current flows at a constant speed - essentially forever. Can then one use the quantum vacuum as a clock? This is a fascinating dispute in condensed-matter physics \cite{wilczek}, formulated as the problem of existence of time crystals.  A time crystal, by analogy with a crystal in space, is a system that displays a time-regularity under measurement, while being in the ground (vacuum) state. These systems might not exist in the form originally proposed \cite{not}, but the research into this new concept will probably bring up unexpected connections between time, the quantum vacuum, and the concept of spontaneously broken symmetry.

Then, if there is an energy (the zero-point energy) associated with empty space, it follows {\it via} the special theory of relativity that this energy should correspond to an inertial mass. By the principle of equivalence of the general theory of relativity, inertial mass is identical with the gravitational mass. Thus, empty space must gravitate.  So, how much does empty space weigh? This question brings us to the frontiers of our knowledge of vacuum - the famous problem of the cosmological constant, a problem that Einstein was wrestling with, and which is still an open issue in modern cosmology \cite{rugh,volovik_myths}.

Finally, although we cannot locally extract the zero-point energy of the vacuum fluctuations, the vacuum state of a field can be used to transfer energy from one place to another by using only information. This protocol has been called {\it quantum energy teleportation} \cite{hotta} and uses the fact that different spatial regions of a quantum field in the ground state are entangled. It then becomes possible to extract locally energy from the vacuum by making a measurement in one place, then communicating the result to an experimentalist in a spatially remote region,  who would be able then to extract energy by making an appropriate (depending on the result communicated)
measurement on her or his local vacuum.

All of the above suggest that the vacuum is the primordial essence, the {\it ousia} from which everything came into existence. Some models suggests that even the spacetime can be seen as an emergent structure. So does Nature try to tell us something about the grand metaphysical question - why there is something rather than nothing - but what exactly \cite{albert}? Does vacuum play the crucial role in the coming into existence of the Universe as we know it \cite{boi,dispute}?

\section{Conclusions}

To conclude, I describe the concept of quantum vacuum in close relation with the latest experimental results that show how particles can be generated by processes such as the dynamical Casimir effect. I then explore the Newtonian-physics assumptions behind the concept of property and show how these are to be modified by relativity and especially by quantum physics. Quantum physics allows for the vacuum state to have an intrinsic structure that provides the ``possibility grid'' for events, or for entities that we can call real with full confidence. Potentialities are thus actualized as properties when the vacuum is disturbed or measured in specific ways.

The emergence of properties by this mechanism sheds new light onto the intricate relation between the quantum and the classical, but does not solve the deep clash between these worlds. Fundamentally, it is perhaps the concept of separation that would need revising. Vacuum itself is possible because one can separate things from one region of space into another. In quantum physics we have the separation between the object under study and the observer (the measurement apparatus). The object under study is quantum while the observer is classical - thus each of them is thought of as obeying a different dynamics, the unitary quantum evolution of the wave function for the object, and the classical equations of motion for the observer. The interaction between the two collapses the wavefunction, resulting in a nonunitary evolution of the object.
This separation does not exist as such in general relativity - there everything, that is, both the object under study and the observer are part of the same dynamical equation: they experience the curvature of spacetime and, by virtue of having mass, they generate the gravitational field themselves. Yet at the same time, quantum physics allows for the existence of entanglement between objects that are localized at different places in space, a feature that seems difficult to accommodate with the theory of relativity. Merging quantum theory with gravitation will therefore most likely require drastically new concepts, also from the direction of what `` emptiness'' means. A frontal approach to the problem of merging gravitation and quantum physics - attempting for example to quantize the gravitational field - might not the best way to proceed, since quantum physics assumes (and hides it very well in the formalism) the existence of a spacetime background in which the measuring apparatus is placed. In other words, the distinction between the object to be quantized and this background cannot be maintained when the object is the spacetime itself.

\acknowledgements
I wish to thank Prof. I. P\^{a}rvu for inviting me to contribute to this volume and for many inspirational discussions during my studies at the Department of Philosophy of the University of Bucharest. The title of the third section of this contribution is inspired from his book, {\it Arhitectura existentei} (Humanitas, Bucharest, 1990). Financial support from FQXi and from the Academy of Finland, projects 263457 and the Center of Excellence ``Low Temperature Quantum Phenomena and Devices'' project 250280 is acknowledged.

I am grateful to Grisha Volovik, Gil Jannes, Janne Karim\"aki, Iulian Toader, and Jonathan Bain for comments on the manuscript.



\begin{thebibliography}{99}

\bibitem{grant} E. Grant,  {\it Much ado about nothing: theories of space and vacuum from the Middle Ages to the scientific revolution}, Cambridge University Press (1981).

\bibitem{Descartes} R. Descartes, {\it Principia Philosophiae} (1644), trans. V.R. Miller and R.P. Miller, Reidel,
Dordrecht (1983).

\bibitem{Newton} I. Newton, `Scholium to the Definitions' in {\it Philosophiae Naturalis Principia Mathematica} (1689); trans. A.  Motte (1729), rev. F. Cajori, Berkeley, University of California Press (1934), pp. 6-12.

\bibitem{saundersbrown} S. Saunders and H. R. Brown, `Reflections on the Ether', pp. 28-63, in S. Saunders and H. R. Brown, {\it The philosophy of vacuum}, Clarendon Press, Oxford (1991).

\bibitem{Einstein} A. Einstein, `Relativity and the Problem of Space', in {\it Relativity - The special and the general theory} (1952), 15th
edition, Three Rivers Press, New York (1961).

\bibitem{aaronson} S. Aaronson, {\it Quantum computing since Democritus}, Cambridge University Press, New York (2013), Ch. 9.

\bibitem{amps} I leave aside Hawkings' black hole information paradox and the hotly debated issue of ``firewalls'', recently triggered by A. Almheiri, D. Marolf, J. Polchinski, and J. Sully,  `Black Holes: Complementarity or Firewalls?', J. High Energy Phys. {\bf 02}, 062 (2013) [arXiv:1207.3123 (2012)]; cf. S. L. Braunstein, `Black hole entropy as entropy of
entanglement, or it's curtains for the equivalence principle'
[arXiv:0907.1190v1] published as S. L. Braunstein, S.
Pirandola and K. \.Zyczkowski, `Better Late than Never: Information
Retrieval from Black Holes', Phy. Rev. Lett. {\bf 110}, 101301 (2013), which deserve a separate discussion.

\bibitem{wigner} E. P. Wigner, `Relativistic Invariance and Quantum Phenomena', Rev. Mod. Phys. {\bf 29}, 255 (1957).

\bibitem{popper} K. Popper, {\it The Logic of Scientific Discovery}, Basic Books, New York, (1959).

\bibitem{kuhn} T. S. Kuhn,  {\it The Structure of Scientific Revolutions}, University of Chicago Press, Chicago (1962).

\bibitem{paraoanu} G. S. Paraoanu, `Realism and single-quanta nonlocality', {\it Found. Phys.} {\bf 41}, 734 (2011).

\bibitem{persp} G. S. Paraoanu, `Quantum computing: theoretical possibility versus practical possibility', {\it Phys. Perspect.} {\bf 13}, 359 (2011).

\bibitem{rovelli} D. Colosi and C. Rovelli,  `What is a particle?', {\it Class. Quant. Grav.} {\bf 26}, 025002 (2009).

\bibitem{redhead} M. Redhead, `The Vacuum in Relativistic Quantum Field Theory', {\it PSA: Proceedings of the Biennial Meeting of the Philosophy of Science Association} {\bf 1994}, Volume Two: Symposia and Invited Papers, 77 (1994).

\bibitem{sciama} D. W. Sciama, `The Physical Significance of the Vacuum State of a Quantum Field', pp. 136-158, in
S. Saunders and H. R. Brown, {\it The philosophy of vacuum}, Clarendon Press, Oxford (1991).

\bibitem{milonni} P. W. Milonni, {\it The Quantum Vacuum: An Introduction to Quantum Electrodynamics}, Academic Press, San Diego, California, U.S.A. (1994).

\bibitem{alpha} P. Indelicato and A. Karpov, `Theoretical physics: Sizing up atoms', {\it Nature} {\bf 498}, 40 (2013)

\bibitem{nuclear} J. Rafelski, L. P. Fulcher, A. Klein, `Fermions and bosons interacting with arbitrarily strong external fields', {\it Phys. Rep.} {\bf 38}, 227 (1978).

\bibitem{Moore}  G. T. Moore,  `Quantum theory of the electromagnetic
field in a variable-length one-dimensional cavity',
{\it J. Math. Phys.} {\bf 11}, 2679 (1970).

\bibitem{ChalmersNature} C. M. Wilson, G. Johansson, A. Pourkabirian, J. R. Johansson, T. Duty, F. Nori, P. Delsing, `Observation of the dynamical Casimir effect in a superconducting
circuit', {\it Nature} {\bf 479}, 376 (2011).

\bibitem{PNAS} P. L\"ahteenm\"aki, G. S. Paraoanu, J. Hassel, and P. J. Hakonen,
`Dynamical Casimir effect in a Josephson metamaterial', {\it Proc. Natl. Acad. Sci. U.S.A.} {\bf 110}, 4234 (2013).

\bibitem{Volovik} G. Volovik, {\it The Universe in a Helium droplet}, Clarendon Press, International series of monographs on physics, Oxford (2003).

\bibitem{westbrook} J.-C. Jaskula, G. B. Partridge, M. Bonneau, R. Lopes, J. Ruaudel, D. Boiron, C. I. Westbrook, `An acoustic analog to the dynamical Casimir effect in a Bose-Einstein condensate', {\it Phys. Rev. Lett.} {\bf 109}, 220401 (2012).

\bibitem{unruh} W. G. Unruh,  `Notes on black-hole evaporation', {\it Phys. Rev. D} {\bf 14}, 870 (1976).

\bibitem{hawking} S. W. Hawking, `Particle creation by black holes', {\it Commun. Math. Phys.} {\bf 43}, 199 (1975).

\bibitem{nanomech} A. D. O'Connell, M. Hofheinz, M. Ansmann, R. C. Bialczak, M. Lenander, E. Lucero, M. Neeley,
D. Sank, H. Wang, M. Weides, J. Wenner, J. M. Martinis, and A. N. Cleland, `Quantum ground state and single-phonon
control of a mechanical resonator', Nature {\bf 464}, 697 (2010).

\bibitem{heisenberg} W. Heisenberg, {\it Physics and Philosophy - The Revolution in Modern Science}, Penguin Books (2000), Ch. 5.

\bibitem{whitehead} A. N. Whitehead, {\it Process and Reality: An Essay in Cosmology}, corrected edition, edited by D. R. Griffin and D. W. Sherburne,  MacMillan Publishing Co. (1979).

\bibitem{hattich} There exists another interpretation, which associates the vacuum state as defined in algebraic field theory with Whitehead's notion of ``universal underlying activity'', see F.  H\"attich, {\it Quantum processes: A Whiteheadian interpretation of quantum field theory}, Agenda Verlag, M\"unster (2004); see also J. Bain, `Quantum processes: A Whiteheadian interpretation of quantum field theory', {\it Studies in History and Philosophy of
Modern Physics} {\bf 36}, 680 (2005).

\bibitem{pregeometry} D. Meschini, M. Lehto, and J. Piilonen, `Geometry, pregeometry and beyond', {\it Studies in History and Philosophy of
Modern Physics} {\bf 36}, 435 (2005).

\bibitem{boi} See especially L. Boi, {\it The Quantum Vacuum: A Scientific and Philosophical Concept, from Electrodynamics to String Theory and the Geometry of the Microscopic World}, John Hopkins University Press,
    Baltimore, Maryland (2011).

\bibitem{bain} J. Bain, `The emergence of spacetime in condensed matter approaches
to quantum gravity',
{\it Studies in History and Philosophy of Modern Physics} {\bf 44}, 338 (2013).

\bibitem{markopoulou} See A. Hamma, F. Markopoulou, S. Lloyd, F. Caravelli, S. Severini, and K. Markstr\"om, `A quantum Bose-Hubbard model with evolving graph as toy model for emergent spacetime',
{\it Phys. Rev. D} {\bf 81}, 104032 (2010), and references therein.

\bibitem{wilczek} F. Wilczek, ``Quantum Time Crystals'', {\it Phys. Rev. Lett.} {\bf 109}, 160401 (2012).

\bibitem{not} P. Bruno, ``Impossibility of Spontaneously Rotating Time Crystals: A No-Go Theorem'', {\it Phys. Rev. Lett.} {\bf 111}, 070402 (2013).

\bibitem{rugh} S. E. Rugh and H. Zinkernagel, `The quantum vacuum and the cosmological
constant problem', {\it Stud. Hist. Phil. Mod. Phys.} {\bf 33}, 663 (2002).

\bibitem{volovik_myths} G. E. Volovik,  `Vacuum Energy: Myths and Reality', {\it Int. J. Mod. Theor. Phys.}
A {\bf 15}, 1987 (2006).

\bibitem{hotta}  N. Hotta, ``Quantum measurement information as a key to energy extraction from local vacuums'',  {\it Phys. Rev. D} {\bf 78}, 045006 (2008).

\bibitem{albert} D. Z. Albert, `On the Possibility That the Present Quantum State of the Universe is the Vacuum', {\it PSA: Proceedings of the Biennial Meeting of the Philosophy of Science Association} {\bf 1988}, Volume Two: Symposia and Invited Papers, 127 (1988).

\bibitem{dispute} L. M. Krauss,  {\it A Universe from Nothing: Why There Is Something Rather Than Nothing}, Free Press, New York (2012). For the dispute that errupted between Lawrence Krauss and philosopher David Albert, see {\it e.g.} D. Albert, `On the Origin of Everything', {The New York Times} Sunday Book Review, 23 March 2012; L.M. Krauss, `The Consolation of Philosophy', {\it Scientific American} April 27 (2012); Sean Caroll, `A Universe from Nothing?', Cosmic Variance,  Discover Magazine, 28 April (2012), {\it etc.}.






\end{thebibliography}
\end{document}